\documentclass[aps,pra,twocolumn,showpacs,superscriptaddress]{revtex4} 
\usepackage{graphicx}
\usepackage{amsmath}
\usepackage{amssymb}
\usepackage{epstopdf}
\usepackage{amsfonts}
\usepackage{dcolumn}
\usepackage{bigints}
\input{epsf}

\usepackage{dcolumn}
\usepackage{bm}
\usepackage[hidelinks,colorlinks=true,linkcolor=blue,citecolor=blue]{hyperref}
\usepackage[mathlines]{lineno}

\usepackage{braket}
\usepackage{color}
\usepackage{float}

\begin{document}
\title{Dynamical mean-field driven spinor condensate physics beyond
the 
single-mode approximation
}
\author{Jianwen Jie}
\address{Homer L. Dodge Department of Physics and Astronomy,
  The University of Oklahoma,
  440 W. Brooks Street,
  Norman,
Oklahoma 73019, USA}
\address{Center for Quantum Research and Technology,
  The University of Oklahoma,
  440 W. Brooks Street,
  Norman,
Oklahoma 73019, USA}
\address{Shenzhen Institute for Quantum Science and Engineering (SIQSE), Southern University of Science and Technology, Shenzhen, P. R. China.}
\author{Shan Zhong}
\address{Homer L. Dodge Department of Physics and Astronomy,
  The University of Oklahoma,
  440 W. Brooks Street,
  Norman,
Oklahoma 73019, USA}
\address{Center for Quantum Research and Technology,
  The University of Oklahoma,
  440 W. Brooks Street,
  Norman,
Oklahoma 73019, USA}
\author{Qimin Zhang}
\address{Homer L. Dodge Department of Physics and Astronomy,
  The University of Oklahoma,
  440 W. Brooks Street,
  Norman,
Oklahoma 73019, USA}
\address{Center for Quantum Research and Technology,
  The University of Oklahoma,
  440 W. Brooks Street,
  Norman,
Oklahoma 73019, USA}
\author{Isaiah Morgenstern}
\address{Homer L. Dodge Department of Physics and Astronomy,
  The University of Oklahoma,
  440 W. Brooks Street,
  Norman,
Oklahoma 73019, USA}
\address{Center for Quantum Research and Technology,
  The University of Oklahoma,
  440 W. Brooks Street,
  Norman,
Oklahoma 73019, USA}
\author{Hio Giap Ooi}
\address{Homer L. Dodge Department of Physics and Astronomy,
  The University of Oklahoma,
  440 W. Brooks Street,
  Norman,
Oklahoma 73019, USA}
\address{Center for Quantum Research and Technology,
  The University of Oklahoma,
  440 W. Brooks Street,
  Norman,
Oklahoma 73019, USA}
\author{Q. Guan}
\address{Homer L. Dodge Department of Physics and Astronomy,
  The University of Oklahoma,
  440 W. Brooks Street,
  Norman,
Oklahoma 73019, USA}
\address{Center for Quantum Research and Technology,
  The University of Oklahoma,
  440 W. Brooks Street,
  Norman,
Oklahoma 73019, USA}
\address{Department of Physics and Astronomy, Washington State University, Pullman, Washington 99164-2814, USA}
\author{Anita Bhagat}
\address{Homer L. Dodge Department of Physics and Astronomy,
  The University of Oklahoma,
  440 W. Brooks Street,
  Norman,
Oklahoma 73019, USA}
\address{Center for Quantum Research and Technology,
  The University of Oklahoma,
  440 W. Brooks Street,
  Norman,
Oklahoma 73019, USA}
\author{Delaram Nematollahi}
\address{Homer L. Dodge Department of Physics and Astronomy,
  The University of Oklahoma,
  440 W. Brooks Street,
  Norman,
Oklahoma 73019, USA}
\address{Center for Quantum Research and Technology,
  The University of Oklahoma,
  440 W. Brooks Street,
  Norman,
Oklahoma 73019, USA}
\author{A. Schwettmann}
\address{Homer L. Dodge Department of Physics and Astronomy,
  The University of Oklahoma,
  440 W. Brooks Street,
  Norman,
Oklahoma 73019, USA}
\address{Center for Quantum Research and Technology,
  The University of Oklahoma,
  440 W. Brooks Street,
  Norman,
Oklahoma 73019, USA}
\author{D. Blume}
\address{Homer L. Dodge Department of Physics and Astronomy,
  The University of Oklahoma,
  440 W. Brooks Street,
  Norman,
Oklahoma 73019, USA}
\address{Center for Quantum Research and Technology,
  The University of Oklahoma,
  440 W. Brooks Street,
  Norman,
Oklahoma 73019, USA}
\date{\today}

\begin{abstract}
$^{23}$Na spin-1 Bose-Einstein condensates are used to
experimentally demonstrate that mean-field physics beyond the single-mode
approximation can be relevant during the 
non-equilibrium dynamics.
The experimentally observed spin oscillation dynamics and associated dynamical spatial structure
formation confirm theoretical predictions that are derived 
by solving a set of coupled mean-field Gross-Pitaevskii 
equations [J. Jie~{\em{et al.}}, Phys. Rev. A {\bf{102}}, 023324 (2020)].
The experiments rely on microwave dressing 
of the $f=1$ hyperfine states, where
$f$ denotes the total angular momentum of the
$^{23}$Na atom.
The fact that beyond single-mode approximation physics at the mean-field level, i.e., spatial mean-field dynamics
that distinguishes the spatial density profiles associated with different Zeeman levels, can---in certain parameter  
regimes---have a pronounced 
effect on the dynamics when the spin healing length
is comparable to or larger than the size of the  Bose-Einstein condensate
has implications for using Bose-Einstein condensates as models for
quantum phase transitions and spin squeezing studies as well as
for non-linear SU(1,1) interferometers.
\end{abstract}
\maketitle

\section{Introduction}
Spinor Bose-Einstein condensates (BECs)
provide an exciting platform for exploring---among 
other phenomena---the dynamics of a quantum pendulum~\cite{pendulum1},
thermal and 
quantum phase transitions~\cite{spinorreview1,spinorreview2,phasetransition0,phasetransition1,phasetransition2,phasetransition3,phasetransition4,phasetransition5,phasetransition6,phasetransition7}, 
 SU(1,1) interferometers~\cite{interferometer1,interferometer2,interferometer3,interferometer4,interferometer5,interferometer6,interferometer7,interferometer8}, and the interplay of symmetry and interactions~\cite{evrard2021a}.
Compared to a single-component BEC,
the spin degrees of freedom
of spinor BECs lead to rich mean-field and beyond mean-field phases
that are characterized by non-trivial order parameters~\cite{ho1998,spinorreview1,spinorreview2}.
In some instances, the spatial orbitals of the different spinor
components are,  to a good approximation, the same:
While the number of atoms occupying each spinor component may be different,
the shape of the spatial orbital is approximately independent of the spinor 
component~\cite{spinorreview1,spinorreview2,sma1,sma2,sma3,sma4}.
This single-mode regime is
said to be realized when the spin
healing length $\xi_s$ is comparable to or larger than the size $R$ of the 
BEC~\cite{dalfovo}.
If $\xi_s \gtrsim R$,
then the BEC is too small to support a ground or low-energy state that exhibits
long wavelength inhomogeneities of the order of the 
size of the BEC, besides those
that exist due to the finiteness of the BEC. In this case,
the densities of the spinor components all have a maximum at the center
of the BEC
and decrease monotonically
till they are zero at the edge of the cloud.

This work presents experimental
data for a spin-1 BEC, which---in conjunction with simulations based on a set of
coupled mean-field Gross-Pitaevskii 
equations---confirm the existence of an alternative mechanism for the creation
of long wavelength density deformations  (i.e., 
density deformations with characteristic length scale of the size of the BEC).
This dynamical mean-field driven mechanism, which is beyond the
single-mode approximation (SMA), has been 
recently predicted theoretically~\cite{jie2020}.
It is distinct from the quantum fluctuation driven processes discussed
in Refs.~\cite{scherer2010,evrard2021} and also distinct from the moving lattice driven process discussed in Ref.~\cite{shaw2022}.
While our work employs a $^{23}$Na BEC, the effect should also be observable in
other spin-1 BECs as well as higher-spin BECs with $s$-wave
contact interactions.
The presence of non-local potentials such as dipolar interactions or spin-orbit coupling,
which couple different partial waves,
would likely modify the observations and interpretation
thereof.

\begin{widetext}

\begin{figure}
\includegraphics[width=1.0\textwidth]{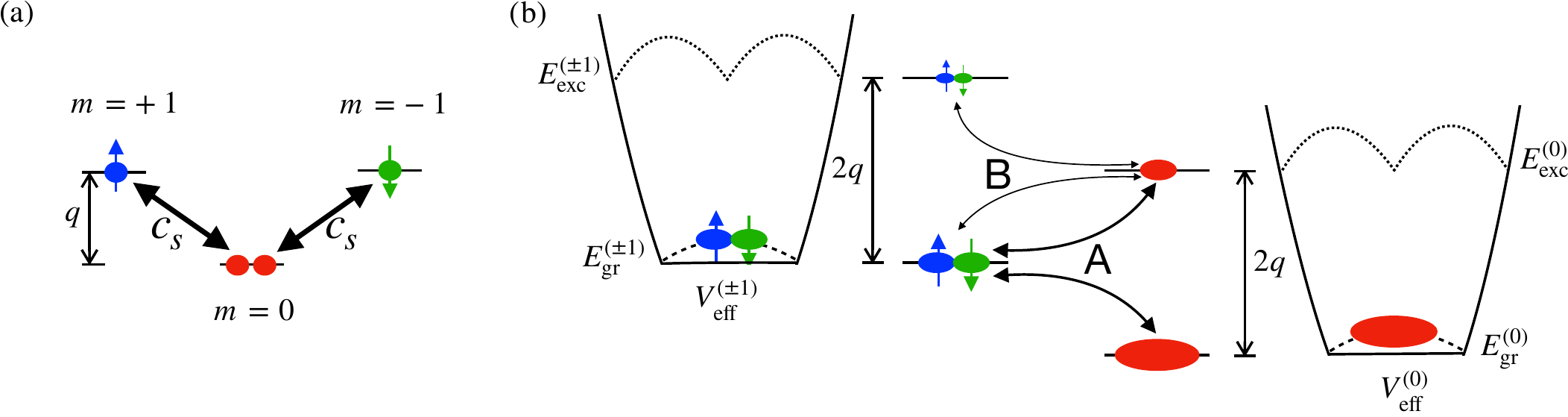}
\caption{Schematic illustration of population changing processes in spin-1 BECs for positive $q$.
(a) ``Standard" spin interaction energy driven process.
The horizontal lines show the single-particle energy levels of the $m=0$ and
$m = \pm 1$ states with single-particle
detuning $q$. The relative shift of the energy levels
is due to the effective quadratic Zeeman shift; the energy contributions
due to the linear Zeeman shift are not shown.
Spin-changing two-body collisions,
characterized by the spin interaction energy $c_s$, lead to population transfer
between the spin-components.
(b) Mean-field driven beyond SMA process.
The effective mean-field potentials $V_{\text{eff}}^{(m)}$
(solid lines; not to scale) felt by the $m= \pm 1$ 
components (left) and $m=0$ component (right) support 
ground and radially excited
states (the corresponding densities are represented schematically by
dashed and dotted lines, respectively). 
If the excitation energy is equal to $2q$, then the
resonant  energy condition facilitates
(A) population transfer from the $m= \pm 1$ 
ground state to the $m=0$ ground and excited states
(and vice versa)  
and 
(B) population transfer from 
the $m=0$ excited state  to the $m=\pm 1$ ground and excited states
(and vice versa). Process (A) (thick curved black arrows), which involves
one excited state, is ``activated" dynamically 
before process (B) (thin curved black arrows), which involves two excited states.
}
\label{fig1}
\end{figure}    

\end{widetext}

The phenomenon described in this paper hinges critically
on the microwave tunability of the
$f=1$ hyperfine energy levels via coupling to the
$f=2$ states~\cite{zhao2014,gerbier2006}; here, 
$f$ denotes the total angular momentum of the atom.
Specifically, a combination of external microwave and magnetic fields is 
used to adjust the single-particle detuning
$q$ between the $m=0$ and $m= \pm 1$ states of the $f=1$ hyperfine manifold.
It is well established that spin-spin interactions,
characterized by the spin interaction energy $c_s$,
are associated with projection quantum number preserving collisions between
two $m=0$ atoms and a pair of $m=\pm 1$ atoms
[see Fig.~\ref{fig1}(a)]. These collisions play an important role in 
quench-induced oscillations of the fractional populations
$\rho_m$ (so-called spin 
oscillations)~\cite{phasetransition3,zhao2014,chang2005,black2007,schmaljohann2004,chang2004,kron2005,kron2006,zibold2016,gerbier2019},
which are---in the SMA framework---governed by the ratios $q/c_s$
and $c_s t / \hbar$, where $t$ denotes the time.
In the beyond-SMA scenario considered in this paper,
the single-particle detuning $q$ is adjusted such that projection quantum
number preserving population transfer is facilitated by ``activating" a long wavelength
excitation.  
The resonance occurs at $q/c_s$ values that are
larger than the critical value at which the spin oscillation
period—predicted within the SMA—diverges~\cite{zhang2005,spinorreview1,spinorreview2}.

Specifically, when $q$ is tuned such 
that $E_{\text{gr}}^{(0)}+E_{\text{exc}}^{(0)}$
is equal to $E_{\text{gr}}^{(+1)} + E_{\text{gr}}^{(-1)}$, 
 the pathway 
 $| m=0,n_{\rho}=0 \rangle + | m=0,n_{\rho}=1 \rangle 
  \leftrightarrow  | m=+1,n_{\rho}=0 \rangle  +  | m=-1,n_{\rho}=0 \rangle$
 becomes resonantly enhanced [see Fig.~\ref{fig1}(b)]~\cite{jie2020}.
 Here, $E_{\text{gr/exc}}^{(m)}$ denotes the energy of
 the ground/excited state (labeled by $n_{\rho}$)
that is supported by the effective mean-field potential associated with the
 $m$th channel.
 The effective potentials have a spatial extent
 that is set by the density interaction energy
 $c_n$,
 thereby supporting an excited state with energy $E_{\text{exc}}^{(m)}$ that sits by an energy
 that is comparable to the
 Thomas-Fermi energy above the ground state
 with energy $E_{\text{gr}}^{(m)}$.
 When the resonance condition is fulfilled, the quench-induced spin oscillation dynamics
 is no longer fully captured by the SMA
 but instead displays,
 as illustrated in this work, oscillations  that
 are characterized by an amplitude and oscillation period that
 change with time; we use the term ``drifting" to refer to this beyond-the-SMA dynamics.
 Since the drifting is captured by the
 coupled Gross-Pitaevskii equations,
 the dynamically induced beyond SMA physics discussed here is 
mean-field in nature; quantum fluctuations are not at play.

The remainder of this paper is organized as follows. Section~\ref{sec_experimental_procedure}
outlines the experimental procedure.
Section~\ref{sec_theory} summarizes the employed mean-field formulation and highlights
mean-field predictions relevant to the experiment.
Section~\ref{sec_experiment} presents and interprets experimental data that evidence
the dynamical emergence of beyond-single-spatial-mode behavior in the 
mean-field regime where the
spin healing length is comparable to or larger than the size of the BEC.
Last, Sec.~\ref{sec_conclusions} summarizes.

\section{Experimental procedure}
\label{sec_experimental_procedure}
Our experiment starts with a nearly pure $^{23}$Na BEC in the
$f=1$, $m=-1$ hyperfine state
in a crossed-beam optical dipole trap.
The trapping potential near the minimum is approximately
harmonic and approximately axially symmetric.
The stronger confinement direction aligns with the direction of gravity.
The center-of-mass sloshing motion, induced either by letting the BEC fall
for a short time before recapturing it or by applying a magnetic field
gradient along the $z$-direction, is used to calibrate the angular frequency
$\omega_z$. 
To measure $\omega_x$ and $\omega_y$, we simultaneously excite
sloshing motions along the $x$- and $y$-directions. 
From the combined motion we deduce that $\omega_x$ and $\omega_y$ are
approximately equal.
In our theory calculations, we set 
$\omega_x=\omega_y=\omega_{\rho}$. 
Calibration measurements yield trap frequencies with an uncertainty of 3~Hz.
While the trap frequencies are stable
for each experimental run, 
variations on the order of up to about $10$~\%
 arise over the course of a measurement 
campaign that lasts around 100~hours due to fluctuations in the laser power and room temperature
(which leads to
changes in the alignment during the course of the day/night).
While the majority of our Gross-Pitaevskii simulations (see below) utilize $\omega_z = 2 \pi \times 246$~Hz and $\omega_{\rho} = 2 \pi \times 140$~Hz,
the dependence of the spin oscillations on $\omega_{\rho}$ is illustrated for select cases.

\begin{figure}
\vspace{-4in}
\includegraphics[width=.7\textwidth]{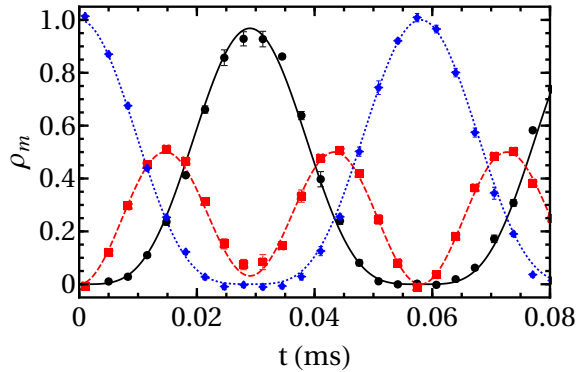}
\caption{State preparation via radio-frequency pulse.
Fractional populations $\rho_m$ of the
$f=1$ hyperfine states are shown as a function of the pulse length.
The symbols show the average of three experimental runs.
Starting with all atoms in the $m=-1$ state (blue triangles),
the radio-frequency pulse with frequency $300$~kHz transfers atoms to the
$m=0$ (red circles) and $m=+1$ (black squares) states.
The lines are the result of a non-interacting three-state
model, which treats the coupling strength of the radio-frequency 
pulse and the magnetic field strength
as fitting parameters.
The fit yields a coupling strength, in units of $h$, of $34.2$~kHz
and $B=0.430$~G.
We estimate the
fluctuations from one initial state preparation
to another to be $0.3$~kHz  
and $0.001$~G  for the coupling strength and magnetic field strength,
respectively. 
}
\label{fig_rfdata}
\end{figure}

The $f=1$ hyperfine levels are split by a constant
magnetic field of $0.430$~G.  The magnetic field corresponds to
a quadratic Zeeman shift of the $f=1$, $m= \pm 1$ levels 
(in units of $h$) by $51.4$~Hz. For the analysis of the data, the linear Zeeman shift is
irrelevant since it can be removed by going to a rotated 
basis~\cite{spinorreview2,sma4}.
To prepare the initial state, we apply a radio-frequency pulse,
which transfers atoms from the
 $m=-1$ state to the 
 $m=0$ and $m=+1$
states (see Fig.~\ref{fig_rfdata}).
The pulse length is chosen such that the fractional
populations of the $m=+1$, $0$, and $-1$
hyperfine states are, to within a few percent, equal to $1/4$, $1/2$, and $1/4$, respectively~\cite{zhao2014,black2007}.

At the end of the radio-frequency pulse 
($t=0$), 
we quench the system by rapidly turning on a microwave field, which dresses
(i.e., shifts) the $m= \pm 1$ hyperfine
states relative to the $m=0$ state.
We parametrize the effective energy shift due to the magnetic field induced quadratic 
Zeeman shift
and the microwave field induced ac-Stark shift by $q$~\cite{zhao2014,gerbier2006}.
Our versatile microwave source~\cite{morgenstern2020}, 
which has the capability to modulate the power
and frequency on fast time scales, provides access to 
a wide range of $q$ values, including positive and negative values~\cite{zhao2014}. 
Throughout this work, we restrict
ourselves to positive $q$; we stress, however, that resonances also exist for negative $q$~\cite{jie2020}.
Our experimental determination of the value of $q$ 
is associated with an uncertainty of
$1$ to $2$~Hz.
The value of $q$ is kept constant
for $0<t< t_{\text{hold}}$. The in-trap dynamics
of the $f=1$ spinor BEC, i.e., the 
quench induced population transfer from 
the $m=0$ state to the $m=+1$ and $-1$ states, 
is then monitored as a function
of $t_{\text{hold}}$.

At $t=t_{\text{hold}}$, the confining potential is
turned off. After $1.5$~ms
of free expansion, a $9$~ms long Stern-Gerlach pulse is applied.
After a total of $10.5$~ms time-of-flight expansion,
destructive absorption imaging
of the $m=1$, $0$, and $-1$ components is performed
in the plane spanned by the unit vectors
$\hat{e}_{xy}=(\hat{x}+\hat{y})/\sqrt{2}$ and $\hat{z}$.
Using standard techniques, we extract the number 
of  
atoms in each of the three spin components
($m=0$ and $\pm 1$) as well as the two-dimensional density.

\section{Theoretical framework and results}
\label{sec_theory}
To describe the spin oscillations that ensue in response to the 
quench at $t=0$ from $q/h=51.4$~Hz for $t<0$ 
to its final value, we employ two different theory frameworks:
the mean-field SMA~\cite{spinorreview1,spinorreview2,sma1,sma2,sma3,sma4} 
and a set of coupled mean-field Gross-Pitaevskii
equations~\cite{spinorreview1,sma2,sma3,sma4,pu2000}.
The former
assumes that the spatial orbitals
of the three spinor components
have an identical shape that is
independent of time. 
The frozen spatial orbital assumption implies a decoupling of
the spatial and spin degrees of freedom.
The spin degrees of freedom are treated at the mean-field
level~\cite{zhang2005}, i.e., the $m=+1$, $0$, and $-1$
components
are characterized by
$\sqrt{\rho_m(t)} \exp [ \mathrm{i} \theta_m(t)]$,
where $\rho_m(t)$
and $\theta_m(t)$
denote the population
and phase of the $m$th component.
Normalization implies $\rho_{+1}(t)+\rho_{0}(t)+\rho_{-1}(t)=1$.
The differential equations that govern
the spin dynamics
[$\rho_0(t)$ and the relative phase $\theta(t)$, where
$\theta(t)$ is defined as $2\theta_0(t)-\theta_{+1}(t)-\theta_{-1}(t)$]
depend on two dimensionless parameters, namely
the ratios $q/c_s$ and $c_s t /h$.
The spin interaction energy $c_s$ is determined by the
spin interaction strength $g_s$,
$g_s=4 \pi \hbar^2(a_2-a_0)/(3M)$ ($M$ denotes the atom mass), and the 
mean spatial density $\overline{n}$ before the
application of the radio-frequency pulse,
$c_s = g_s \overline{n}$. Here,
$a_0$ and $a_2$ denote the two-body scattering lengths in the two-particle angular momentum
channels $0$ and $2$, $a_0=48.91$~$a_B$ and $a_2=54.54$~$a_B$~\cite{knoop2011}
($a_B$ denotes the Bohr radius).
The shape of the spatial orbital, and correspondingly
the mean spatial density $\overline{n}$,
is determined by solving a single-component time-independent
Gross-Pitaevskii equation, which depends on the 
aspect ratio $\lambda$ ($\lambda=\omega_{z}/\omega_{\rho}$)
and the dimensionless interaction strength
$(N-1)g_n / (\hbar \omega_z a_{\text{ho},z}^3)$, where
$g_n=4 \pi \hbar^2(2a_2+a_0)/(3M)$ and $a_{\text{ho},z}^2=\hbar/(M \omega_z)$.
For typical atom numbers and trap frequencies considered in this paper
(i.e., 
$N=2.3 \times 10^{4}$, $\omega_{\rho}= 140$~Hz, and $\lambda = 1.75$),
the associated density interaction energy $c_n$,
$c_n=g_n \overline{n}$, is (in units of $h$)
equal to 
$589$~Hz.
The fact that $c_n$ is $28.1$ times larger than $c_s$
is typically
used to justify the applicability of the
SMA. Within the SMA framework, the spin oscillations are fully periodic (time-independent oscillation period and 
time-independent minimum/maximum amplitude)~\cite{zhang2005}. 

\begin{figure}
\vspace*{0.1in}
\includegraphics[width=.35\textwidth]{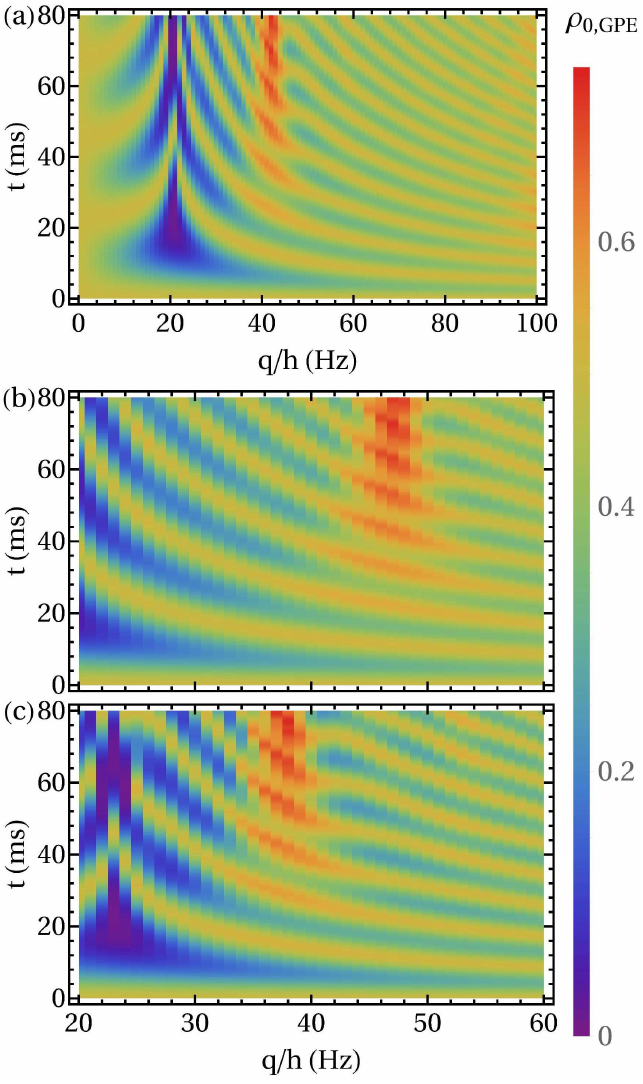}
\caption{Fractional population $\rho_{0,\text{GPE}}(t)$ 
as functions of time $t$ and Zeeman energy $q$ 
for  $\omega_z=2 \pi \times 246$~Hz, $\omega_{\rho}=2 \pi \times 140$~Hz,
and (a) $N=2.3 \times 10^4$, (b) $N=1.7 \times 10^4$, and (c) $N=3.1 \times 10^4$;
the color coding is given by the color bar on the right.
In (a), the oscillation period diverges at $q/h \approx 20$~Hz, in agreement with the mean-field SMA result of $q/h \approx c_s/h=20.9$~Hz.  
Note that the range of $q$ values considered in (a) is larger than
in (b) and (c). 
This paper focuses on the regime where the spin dynamics deviate
from regular oscillatory behaviors; the red color for $q/h \approx 40$~Hz signals drifting.
It can be seen that the $q$-values for which the 
drifting occurs (``fuzzy" red region) move to smaller
values with increasing $N$.}
\label{fig2_theory_new}
\end{figure}

\begin{figure}
\vspace*{0.1in}
\includegraphics[width=.4\textwidth]{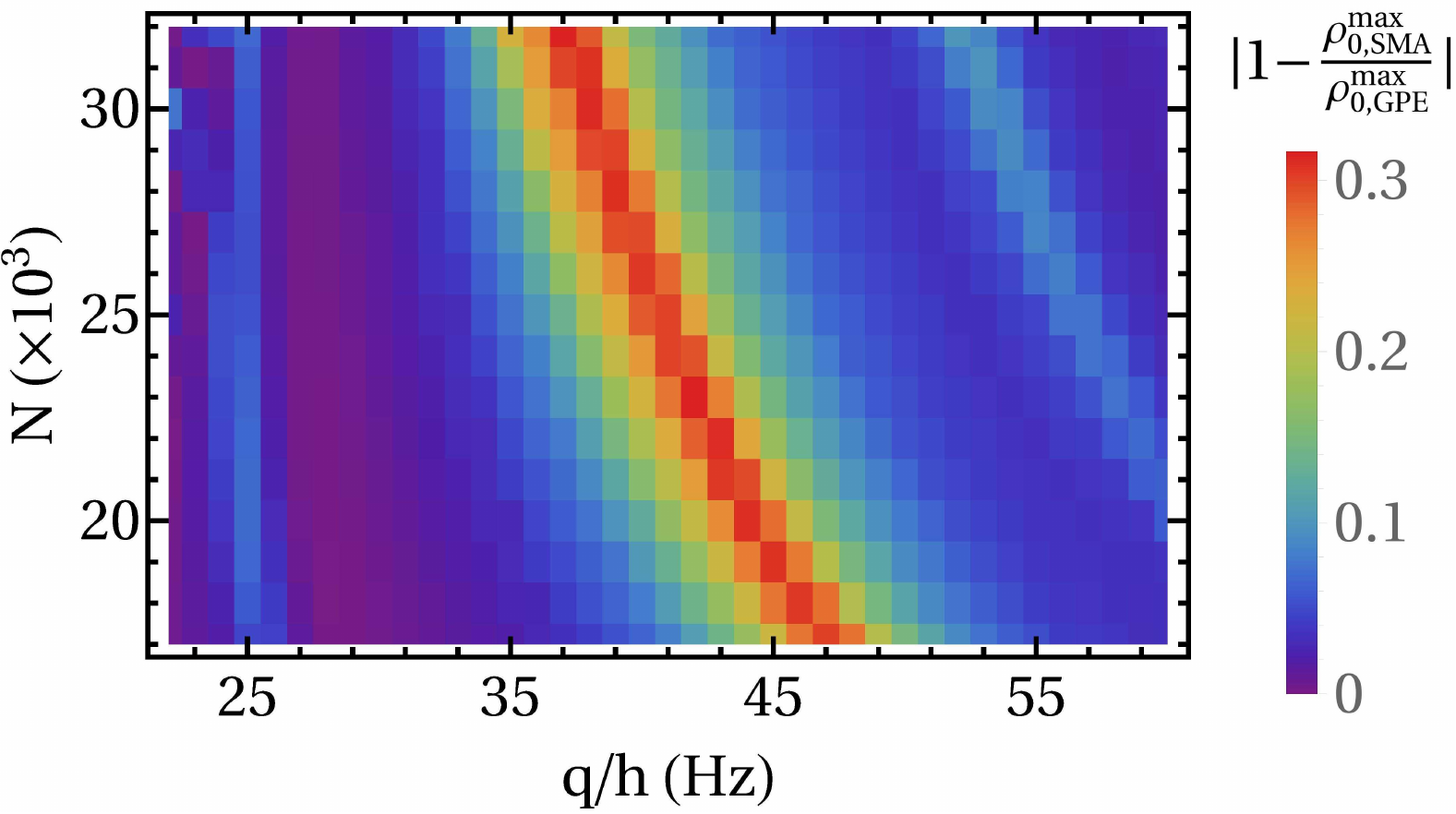}
\caption{Absolute value of the normalized difference $[\rho_{0,\text{GPE}}^{\text{max}}-\rho_{0,\text{SMA}}^{\text{max}}]/\rho_{0,\text{GPE}}^{\text{max}}$,
obtained by considering $0 < t \le 80$~ms, as functions of $N$ and $q/h$ for $\omega_z=2 \pi \times 246$~Hz
and $\omega_{\rho}=2 \pi \times 140$~Hz.
}
\label{fig3_theory_new}
\end{figure}

\begin{figure}
\vspace*{0.1in}
\includegraphics[width=.45\textwidth]{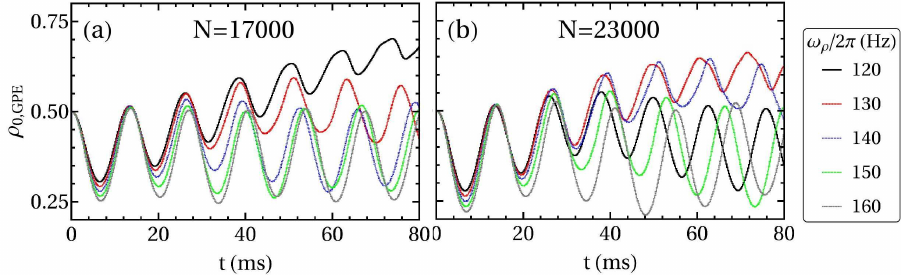}
\caption{Fractional population $\rho_{0,\text{GPE}}(t)$ 
as a function of time $t$ for $q/h=40$~Hz, $\omega_{z}=2 \pi \times 246$~Hz, and
 (a) $N=1.7 \times 10^{4}$ and (b) $N=2.3 \times 10^4$. The different lines
 show results for different 
  $\omega_{\rho}$,
 ranging from $2 \pi \times 120$ to $2 \pi \times 160$~Hz
 (see the legend on the right of the figure).
}
\label{fig4_theory_new}
\end{figure}

To go beyond the SMA,
we solve a set of three coupled time-dependent mean-field Gross-Pitaevskii equations,
which depend on five dimensionless parameters: 
$(N-1)g_n / (\hbar \omega_z a_{\text{ho},z}^3)$,
$g_n/g_s$, $\lambda$,
$q/c_s$, and $t c_s/ \hbar$~\cite{jie2020}.
This framework allows for the coupling of the spin and spatial degrees of freedom,
which---in the regime where the SMA breaks down---can
lead
to modifications of the spin oscillations.
In particular,
previous theory work~\cite{jie2020} predicted that the 
interplay between these degrees of freedom induces, for certain parameter
combinations, a resonance-like effect that leads to drifting, i.e., spin oscillations 
whose oscillation amplitude, frequency, and mean value are not---as predicted by the SMA---constant in time.
Figures~\ref{fig2_theory_new}-\ref{fig4_theory_new} show examples of this behavior.

The physical picture behind the drifting is
illustrated in Fig.~\ref{fig1}.
Within the coupled Gross-Pitaevskii equation framework,
the $m$th spinor component feels an effective time-dependent
mean-field potential that is created by its own spinor wave function as well as 
the spinor wave functions of the other components.
Neglecting some small terms so that the effective potentials depend only on the densities
of the spinor components and treating the time as an adiabatic
parameter~\cite{jie2020}, the effective potential $V_{\text{eff}}^{(m)}(\vec{r},t)$
of the $m$th component supports a 
ground state with energy $E_{\text{gr}}^{(m)}$
and excited states with energies $E_{\text{exc},j}^{(m)}$ at each time.
Specializing to positive $q$, 
a resonance condition is realized when
 $E_{\text{gr}}^{(+1)}+E_{\text{gr}}^{(-1)}$ is equal to
$E_{\text{gr}}^{(0)}+E_{\text{exc},j}^{(0)}$, i.e.,
when a pair of $m=\pm 1$ atoms is energetically degenerate with two $m=0$ atoms,
one in the ground and one in the excited state of the effective potential felt by
the $m=0$ component.
The energetic degeneracy enhances projection quantum number
preserving population transfer between the $m=0$ and $m=\pm 1$ modes (and vice versa).
Since the excited state associated with the energy
$E_{\text{exc},1}^{(0)}$ 
has a ``wavelength" or ``density modulation"
that is of the order of
the size of the BEC,
the considered $m=0 \leftrightarrow m=\pm 1$ population transfer mechanism
leads to dynamically induced density deformations of the spinor components;
for $j \ge 2$, the associated density deformation is characterized by a smaller 
length scale.
The competition between the 
dimensionless energy
scales $q/c_s$ and $[E_{\text{gr}}^{(0)}-E_{\text{exc},j}^{(0)}]/c_s$
triggers the drifting of
the spin oscillations.
For fixed $(N-1)g_n / (\hbar \omega_z a_{\text{ho},z}^3)$,
$g_n/g_s$, and $\lambda$, the single-particle detuning $q$ provides a knob for 
tuning the spinor BEC into and out of resonance.

Figure~\ref{fig2_theory_new} shows the fractional population $\rho_0$,
obtained by evolving the initial state using the time-dependent three-component Gross-Pitaevskii equation for 
various $q/h$ for a $^{23}$Na BEC consisting of (a) $N=2.3 \times 10^4$, (b) $N=1.7 \times 10^4$, and (c) $N=3.1 \times 10^4$ particles. 
  For $N=2.3 \times 10^4$ [Fig.~\ref{fig2_theory_new}(a)], the divergence of the oscillation
period at $q^*/h \approx 20$~Hz---which is associated with a separatrix in classical phase space---is clearly visible and well described by the 
mean-field spin model. If the SMA was valid, the
spacing of the  ``green" and  ``blue" stripes would be decreasing monotonically as one moves 
away from the divergence. While this holds for $q< q^*$, ``irregularities" are observed for $q/h \approx 40$~Hz;
these irregularities are indicative of the drifting that is caused by a resonance (see the discussion in the context of Fig.~\ref{fig1} above).
Figures~\ref{fig2_theory_new}(b)-\ref{fig2_theory_new}(c), which show the dynamics of the
spin oscillations for two other $N$ (using a smaller $q$-region), demonstrate that the irregularities depend quite sensitively
on the  particle number. The 
sensitivity of the drifting on the particle number is an important consideration when 
interpreting the experimental data (see Sec.~\ref{sec_experiment}).

To quantify the deviations between the fractional populations obtained by the mean-field Gross-Pitaevskii framework [$\rho_{0,\text{GPE}}(t)$]
and the SMA-based mean-field spin model [$\rho_{0,\text{SMA}}(t)$; time-independent oscillation period and maximum/minimum amplitude], Fig.~\ref{fig3_theory_new}
shows the absolute value of the normalized difference between the maximum $\rho_{0,\text{GPE}}^{\text{max}}$ of $\rho_{0,\text{GPE}}(t)$  
and the maximum $\rho_{0,\text{SMA}}^{\text{max}}$ of $\rho_{0,\text{SMA}}(t)$, calculated using fractional population data for $0<t \le 80$~ms,
as functions of $N$ and $q$. The quantity $|\rho_{0,\text{GPE}}^{\text{max}}-\rho_{0,\text{SMA}}^{\text{max}}|/\rho_{0,\text{GPE}}^{\text{max}}$  is obtained for the same trap frequencies as those 
considered in Fig.~\ref{fig2_theory_new}. 
A larger value of $|\rho_{0,\text{GPE}}^{\text{max}}-\rho_{0,\text{SMA}}^{\text{max}}|/\rho_{0,\text{GPE}}^{\text{max}}$ signals larger drifting.
Figure~\ref{fig3_theory_new} shows that the drifting depends sensitively
on both $N$, which unavoidably fluctuates in experiment, and $q$, which can be tuned via microwave dressing.

Motivated by the fact the experimental trap frequencies can change
by up to 
about 10~\% 
over the course of a day (see Sec.~\ref{sec_experimental_procedure}), Fig.~\ref{fig4_theory_new} illustrates the dependence of the spin oscillations 
on the angular trapping frequency $\omega_{\rho}$ along the $\rho$-direction.
For both $N$ considered, the deviations 
between the fractional populations for different trap frequencies but otherwise identical parameters initially  increases with time ($t \lesssim 30$~ms).
For some of the parameter combinations [see, e.g., the red line for $2 \pi \times \omega_{\rho}=130$~Hz in Fig.~\ref{fig4_theory_new}(a) and the black line for $2 \pi \times \omega_{\rho}=120$~Hz in Fig.~\ref{fig4_theory_new}(b)], the upward drift slows after a finite number of oscillations, followed by a downward drift; this behavior
is indicative of a competition of two energy scales, namely the spin and the density interaction energies.
Looking ahead to the interpretation of the experimental data, a key message of Fig.~\ref{fig4_theory_new} is that the spin oscillation dynamics depends more strongly on the trap frequencies in the vicinity of the resonance than away from the resonance, i.e., the amount of drifting depends---when it occurs---sensitively on $\omega_{\rho}$.

\section{Experimental results}
\label{sec_experiment}

\begin{figure}
\vspace*{0.1in}
\includegraphics[width=.45\textwidth]{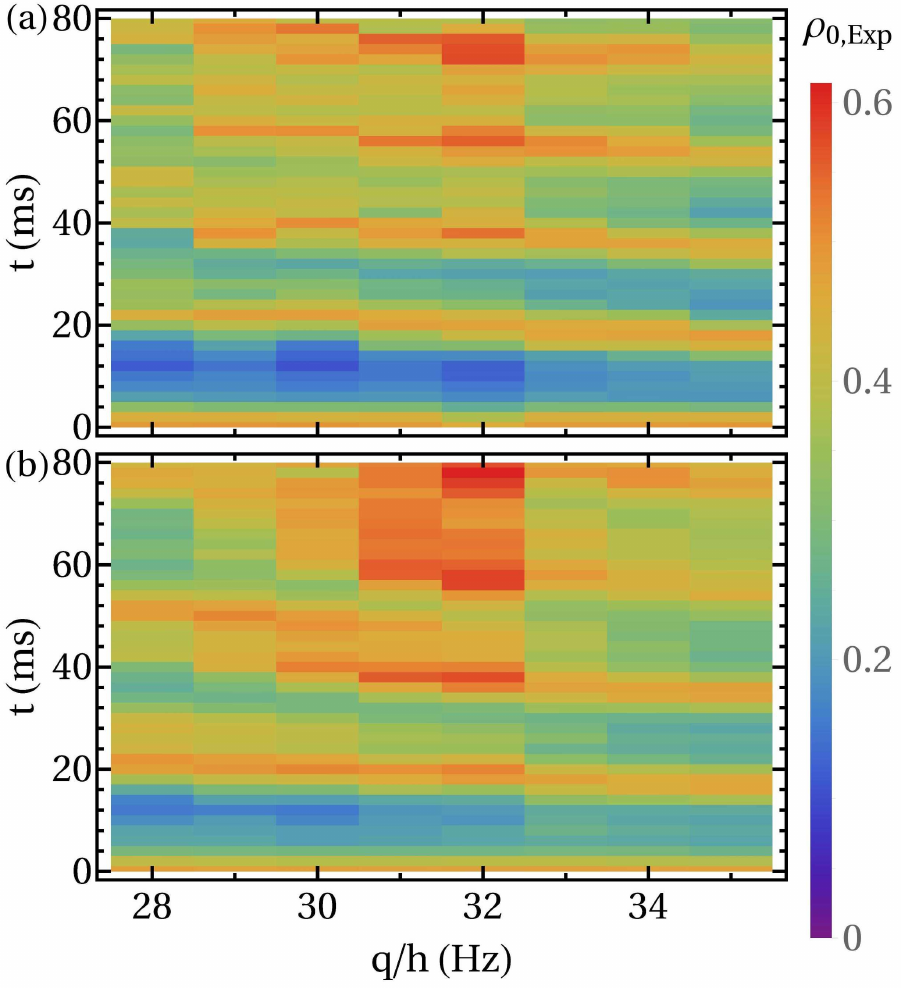}
\caption{Fractional population $\rho_{0,\text{Exp}}$ as functions of the hold time $t=t_{\text{hold}}$ and the Zeeman energy $q/h$ determined in two different
experimental campaigns.
The trap frequencies are (a) 
$(\omega_x, \omega_y, \omega_z)=2\pi\times(147,132,246)$~Hz
and (b) 
$(\omega_x, \omega_y, \omega_z)=2\pi\times(140,122,255)$~Hz.   The 
  red
  regions are interpreted as signatures of the drifting, in qualitative
agreement with what is predicted by mean-field Gross-Pitaevskii simulations (see Fig.~\ref{fig2_theory_new}).}
\label{fig2_exp_new}
\end{figure} 

This section presents experimental data that confirm the dynamically induced beyond SMA
spin-oscillation dynamics.
Analogous to Fig.~\ref{fig2_theory_new},
Figs.~\ref{fig2_exp_new}(a) and \ref{fig2_exp_new}(b) show the fractional population
$\rho_0(t)$ of the $f=1$, $m=0$ hyperfine state as functions of the hold time $t$ and $q$ for two separate data campaigns,
corresponding to somewhat different mean atom numbers and trap frequencies.
The two data campaigns used,  several
months apart, the same apparatus [the data from Fig.~\ref{fig2_exp_new}(a) are also shown in the Supplemental Material~\cite{supplement}].
The 
data sets
shown in Figs.~\ref{fig2_exp_new}(a) and \ref{fig2_exp_new}(b) are  characterized by mean atom numbers of 
$2.3 \times 10^4$ and $2.7 \times 10^4$,
respectively, with BECs ranging from about
$N=1.5 \times 10^4$ to $N=3.2 \times 10^4$. 
The atom number distributions follow Gaussians with standard deviations 
 of $\sigma_N=1.7 \times 10^3$ and $1.8 \times 10^{3}$ in
Figs.~\ref{fig2_exp_new}(a) and \ref{fig2_exp_new}(b), respectively.
A BEC with $N=2.4 \times 10^4$, $\omega_z=2 \pi \times 246$~Hz, and $\omega_{\rho}=2 \pi \times 140$~Hz, e.g., 
corresponds to Thomas-Fermi radii along the $\rho$-
and $z$-directions
of $R_{\text{TF},\rho} \approx 7.06$~$\mu$m and
$R_{\text{TF},z} \approx 0.57 R_{\text{TF},\rho}$, respectively.
For comparison, the spin healing length $\xi_s$,
$\xi_s=\hbar/\sqrt{2M |c_s|}$,
is about
$3.20$~$\mu$m, i.e., the spin healing
length is comparable to the Thomas-Fermi radii in the $\rho$- and
$z$-directions. 
In Fig.~\ref{fig2_exp_new},
the 
hold time and $q$-value are varied in steps of 2~ms and $1$~Hz, respectively.
Each experimental data point is the average of 10~measurements. Drifting can be seen in
both data sets for $q/h$ values around $31-32$~Hz, in qualitative agreement with the theoretical mean-field Gross-Pitaevskii simulations.

\begin{figure}
\includegraphics[width=0.3\textwidth]{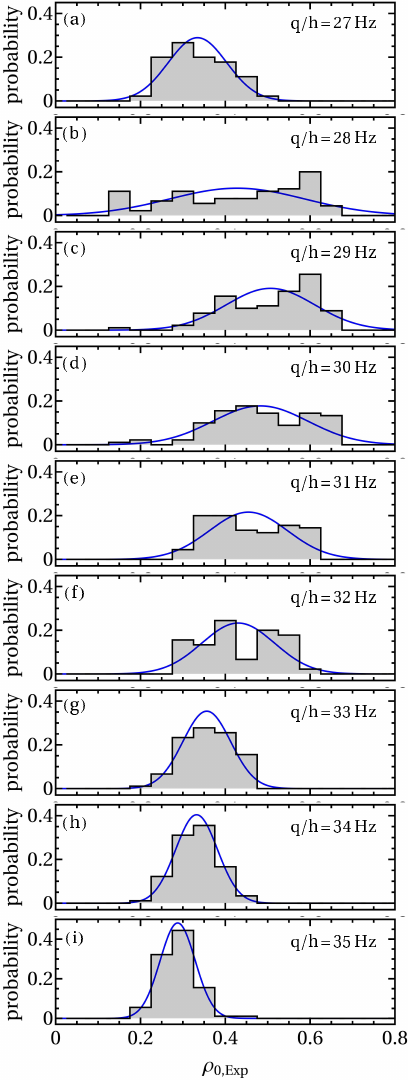}
\caption{Distribution of
fractional population $\rho_{0,\text{Exp}}$ at $t=t_{\text{hold}}=60$~ms for $(\omega_x, \omega_y, \omega_z)=2\pi\times(147,132,246)$~Hz 
for (a) $q/h=27$~Hz to (i) $q/h=35$~Hz. 
The blue lines show Gaussian distributions, using the mean value and standard deviation of the experimentally measured $\rho_{0,\text{Exp}}$.
}
\label{fig_experiment_rho0_histo}
\end{figure}

To map out the resonance in more detail, the gray histograms in Fig.~\ref{fig_experiment_rho0_histo} show the distribution of the fractional population $\rho_{0,\text{Exp}}(t)$ at
$t=t_{\text{hold}}=60$~ms for nine $q/h$-values; for each $q/h$, the experiment is repeated 90 times.
The data are for the same conditions (i.e., same mean particle number and trap frequencies) as 
Fig.~\ref{fig2_exp_new}(a). 
The blue solid lines show normalized Gaussian distributions that are obtained from the mean value and standard deviation of the experimental data. 
It can be seen that $\rho_{0,\text{Exp}}$
(i.e., the fractional population at which the blue lines take their maximum) changes---as expected---smoothly with $q$: it increases monotonically for $q/h=27-29$~Hz and subsequently decreases monotonically for $q/h=29-35$~Hz. The shape of the gray histograms,   in contrast, varies  intricately with $q/h$.  
The distributions for $q/h=27$~Hz and $q/h \ge 33$~Hz are approximately Gaussian and comparatively narrow; for the other $q/h$-values, the distributions are less well described by a Gaussian distribution (fits, not shown, result in larger $\chi^2$) and comparatively broad. 
The non-Gaussian behavior near $q/h=31$~Hz is attributed to the resonance, which triggers the drifting that is---as evidenced by our mean-field Gross-Pitaevskii simulations shown in Fig.~\ref{fig3_theory_new}---associated with a comparatively strong sensitivity to the particle number. As a consequence, repeated experiments with a Gaussian atom number distribution lead to a non-Gaussian 
distribution of the fractional population in the $m=0$ component.

If the system was described accurately by the SMA, the maximum
of the densities $n_{m}(\vec{r},t)$ would always be located at 
$(\rho,z)=(0,0)$, where $\rho^2$ is equal to $x^2+y^2$.
However,
Gross-Pitaevskii simulations for 
an axially-symmetric trap and 
axially-symmetric $m$-dependent mean-field 
orbitals
show that the beyond SMA spin
oscillation dynamics are associated with density deformations
(peak densities that are located at $\rho \ne 0$),
which develop dynamically with increasing time $t=t_{\text{hold}}$~\cite{jie2020}.
These deformations oscillate back and forth between
the $m=0$ and $m= \pm 1$ components. 
While the density deformations are most pronounced on resonance, they also occur for 
$q/h$-values below and above the resonance.

\begin{figure*}
\includegraphics[width=0.9\textwidth]{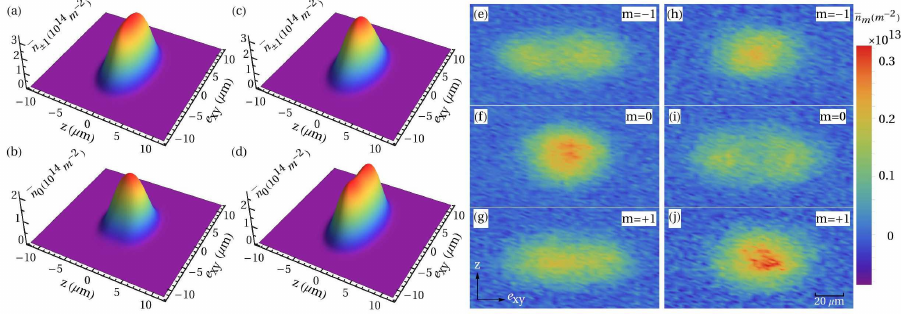}
\caption{Theoretical and experimental spatial integrated two-dimensional densities $\overline{n}_{m}$ [defined through
$\overline{n}_m(e_{xy},z,t)=\int n_m(\vec{r},t) d \eta$, where
$\eta=(x+y)/\sqrt{2}$]
for $q/h=31$~Hz.
(a)-(d) Theoretical densities, obtained by solving the coupled
Gross-Pitaevskii equations for $N=4 \times 10^4$, 
$\omega_{\rho}= 2 \pi \times 140$~Hz, and $\omega_z=2 \pi \times 246$~Hz; (a) $m=\pm1$ and $t=t_{\text{hold}}=50$~ms,
(b) $m=0$ and $t=t_{\text{hold}}=50$~ms,
(c) $m=\pm1$ and $t=t_{\text{hold}}=58$~ms, and
(d) $m=0$ and $t=t_{\text{hold}}=58$~ms.
The normalization is 
$\sum_m \int n_m(\vec{r},t) d\vec{r}=N$.
The time-of-flight sequence is not included in the simulations.
(e)-(j)
Experimental two-dimensional  images 
for $10.5$~ms time-of-flight
expansion; 
(e) $m=-1$ and $t=t_{\text{hold}}=50$~ms,
(f) $m=0$ and $t=t_{\text{hold}}=50$~ms,
(g) $m=+1$ and $t=t_{\text{hold}}=50$~ms,
(h) $m=-1$ and $t=t_{\text{hold}}=58$~ms,
(i) $m=0$ and $t=t_{\text{hold}}=58$~ms,
(j) $m=+1$ and $t=t_{\text{hold}}=58$~ms.
The particle numbers $N$ are $2.6 \times 10^4$ and $2.2 \times 10^4$ for
(e)-(g) and (h)-(j), respectively.
 To aid with the visualization,
the three images were centered individually.
The side bar on the right defines the color code for the experimental images shown in (e)-(j).}
\label{fig_spinoscillation}
\end{figure*}

Figures~\ref{fig_spinoscillation}(a)-\ref{fig_spinoscillation}(d)
show integrated two-dimensional Gross-Pitaevskii component densities $\overline{n}_m(e_{xy},z,t)$ as functions 
of  $z$ and $e_{xy}=(x+y)/\sqrt{2}$
(this is the same representation as employed in
the experimental imaging system)
 for $q/h=31$~Hz,
$\omega_{\rho}= 2 \pi \times 140$~Hz, $\omega_z= 2 \pi \times 246$~Hz,
and
$N=4 \times 10^4$ at
two times, namely $t=50$~ms 
[Figs.~\ref{fig_spinoscillation}(a)-\ref{fig_spinoscillation}(b)]
and $t=58$~ms 
[Figs.~\ref{fig_spinoscillation}(c)-\ref{fig_spinoscillation}(d)]. For this particle number, the system is  close to resonance, as can be seen by extrapolating the simulation results shown in
Fig.~\ref{fig3_theory_new} to larger $N$.
The two images at the top left
show the $m= \pm 1$ densities while the two images at the bottom left
show the 
$m=0$ density.
Since the $(e_{xy},z)$-representation is ``inconsistent" with the axial
symmetry of the system, the density deformations are being partially averaged over.
They lead to an elongation along the $e_{xy}$-direction of the $m=\pm1$ density at $t=50$~ms 
[Fig.~\ref{fig_spinoscillation}(a)] and a double-peak structure of the $m=0$ density at $t=58$~ms [Fig.~\ref{fig_spinoscillation}(d)].

The experimental images shown in 
Figs.~\ref{fig_spinoscillation}(e)-\ref{fig_spinoscillation}(g)
are for
$t=t_{\text{hold}}=50$~ms
while those 
shown in 
Figs.~\ref{fig_spinoscillation}(h)-\ref{fig_spinoscillation}(j) are for $t=t_{\text{hold}}=58$~ms; they
correspond to the same $q/h$-value as the theory calculations but 
 smaller atom number.
It can be seen that the experimental data are in qualitative agreement with the Gross-Pitaevskii simulation results, thereby confirming the beyond SMA dynamics.
Gross-Pitaevskii simulations for the same atom numbers as measured experimentally show significantly smaller density deformations. This is attributed to multiple effects.
The 
experimental set-up breaks the axial symmetry, which is assumed to hold strictly in the theory calculations, weakly. Moreover,  the experimental data may be impacted by small trap frequency variations.
Our simulations show that the resonance position and shape depend---due to the intricate interplay between the kinetic and potential energy contributions---sensitively on the exact trap parameters and atom number
(see Figs.~\ref{fig2_theory_new}-\ref{fig4_theory_new}), rendering fully quantitative side-by-side comparisons of theory and experiment challenging.

\section{Conclusions}
\label{sec_conclusions}
This paper presented theory predictions and experimental data for a sodium spinor condensate
that confirm the existence of a dynamically-induced
mean-field driven resonance mechanism that is not captured by
the SMA. The physical picture behind the resonance mechanism is quite simple:
When the density and spin interaction strengths are such that
the effective mean-field potentials support an excited
state that leads to an energetic degeneracy,
population transfer between the $m=0$ and $m= \pm 1$ modes 
is enhanced. For a fixed
single-particle detuning $q$, the mean-field parameters can
be adjusted by, e.g., changing the
particle number or trap frequencies.
This population transfer mechanism is distinctly different from the
``usual" collision induced population transfer in which the 
spin-changing two-body collision term ``triggers"
the transfer of population. This process is, in the case where the density interaction
energy is much larger than the spin interaction energy, 
captured by the SMA.
The resonance mechanism studied in this paper, in contrast, 
is not captured by the SMA since it 
leads to the dynamical occupation of excited spatial modes. While the experimental observations
reported here are for a  spin-1 $^{23}$Na BEC, the same mechanism exists---according to the
theory---in spin-1 $^{87}$Rb BECs, which are characterized by a much larger
density-to-spin-interaction-energy ratio.

The results presented in this paper are of relevance to a broad range of
dynamical studies involving spinor BECs. Spinor BECs have, e.g., been used
to study quench-induced dynamical quantum phase transitions, which are supported by
the quantum spin Hamiltonian that is derived by treating the spatial degrees of freedom
within the SMA. The quantum spin Hamiltonian also forms the
starting point 
of spin squeezing studies and interferometer protocols. 
The work presented in this paper shows that attention needs to
be paid to the mean-field parameters 
to ensure that the SMA provides a faithful description.
The dynamically induced transfer of population to excited modes, which can be controlled
by adjusting the single-particle detuning via microwave dressing, provides a 
novel route for studying the coupling between the spin and spatial degrees of freedom.

\section*{Acknowledgement}
\label{acknowledgement}
We gratefully acknowledge support
by the National Science Foundation through
grant numbers
PHY-1806259,
PHY-2110158,
and PHY-1846965 (CAREER), and from the Department of Defense through AFOSR Grant No. FA9550-20-1-0071. 
JJ is supported by the National Natural Science Foundation of China under Grant No. 12104210 and the China Postdoctoral Science Foundation under Grant No. 2022M711496.
This work used  the OU
Supercomputing Center for Education and Research
(OSCER) at the University of Oklahoma (OU).

\end{document}